\documentclass[%
 aip,
 rsi,
 amsmath,amssymb,twocolumn,
 reprint,%
]{revtex4-1}
\usepackage{graphicx}
\usepackage{dcolumn}
\usepackage{bm}
\usepackage[mathlines]{lineno}

\usepackage[utf8]{inputenc}
\usepackage[T1]{fontenc}
\usepackage{mathptmx}

\draft 

\begin{document}

\preprint{AIP/123-QED}

\title[Development and characterization of high-frequency sources for  supersonic beams of fluorine radicals]{Development and characterization of high-frequency sources for  supersonic beams of fluorine radicals} 

\author{Patrik Stra\v{n}\'ak}

\author{Ludger Ploenes}

\affiliation{Department of Chemistry, University of Basel, Klingelbergstrasse 80, 4056 Basel, Switzerland}%

\author{Simon Hofs\"ass}
\altaffiliation{Present address: Fritz-Haber-Institut der Max-Planck-Gesellschaft, Faradayweg 4-6, 14195 Berlin, Germany}
\affiliation{Institute of Physics, University of Freiburg, Hermann-Herder-Stra{\ss}e 3, 79104 Freiburg, Germany}

\author{Katrin Dulitz}
\affiliation{Institute of Physics, University of Freiburg, Hermann-Herder-Stra{\ss}e 3, 79104 Freiburg, Germany}

\author{Frank Stienkemeier}
\affiliation{Institute of Physics, University of Freiburg, Hermann-Herder-Stra{\ss}e 3, 79104 Freiburg, Germany}

\author{Stefan Willitsch}
 \email[]{Author to whom correspondence should be addressed. Electronic address: stefan.willitsch@unibas.ch}
\affiliation{Department of Chemistry, University of Basel, Klingelbergstrasse 80, 4056 Basel, Switzerland}%

\date{\today}

\begin{abstract}
We present and compare two high-pressure, high-frequency electric-discharge sources for the generation of supersonic beams of fluorine radicals. The sources are based on dielectric-barrier-discharge (DBD) and plate-discharge units attached to a pulsed solenoid valve. The corrosion-resistant discharge sources were operated with fluorine gas seeded in helium up to backing pressures as high as 30 bar. We employed a (3+1) resonance-enhanced multiphoton ionization combined with velocity-map imaging for the optimization, characterization and comparison of the fluorine beams. Additionally, universal femtosecond-laser-ionization detection was used for the characterization of the discharge sources at experimental repetition rates up to 200 Hz. Our results show that the plate discharge is more efficient in F$_{2}$ dissociation than the DBD by a factor of 8-9, whereas the DBD produces internally colder fluorine radicals. 
\end{abstract}


\maketitle 

\section{Introduction}

As the element with the highest electronegativity, fluorine is reactive with a wide range of chemical compounds. Due to its high reactivity, atomic fluorine is often used in reaction-dynamics studies of fast radical reactions. For instance, recent studies show its significance in revealing dynamical resonances and resonance-induced quantum tunneling in the reaction F+H$_{2}$. \cite{yang08a,yang19a} The mechanisms and dynamics of reactions with fluorine radicals in the gas phase are often studied with crossed-molecular-beam (CMB) techniques which rely on the generation of cold, supersonic beams with high densities.

In early crossed-beam experiments, beams of fluorine radicals were produced by thermal dissociation or microwave discharge of F$_2$ or other fluorine-containing precursors.\cite{sung77a,becker78a,nazar81a} In order to narrow the large spread of the beam velocity inherent in these approaches, velocity selectors were used which, however, also decreased the number densities of radicals and, therefore, reduced the advantage of an efficient production of F radicals afforded by these techniques. \cite{neumark85a,aquilanti90a} Nesbitt and co-workers generated fluorine radicals by implementing a plate-discharge source at the orifice of a pulsed-gas nozzle.\cite{chapman98a,zolot07a} Yang and co-workers further improved the discharge efficiency by introducing a double-stage pulsed discharge source, the stability of which was enhanced by a pre-ionization stage in the discharge. \cite{ren06a} Regular maintenance of the valves was a necessity due to constant flow of highly corrosive gas of molecular fluorine through the moving mechanism of the plunger. Liu and co-workers addressed this problem by separating the actuation mechanism of the plunger from the gas flow by modifying a pulsed piezoelectric-disk valve resulting in increased operation times.\cite{dong00a} Typically, the repetition rates of previous pulsed fluorine-radical sources were rather low, reaching up to 20 Hz. An increase of the repetition rate was recently achieved by the development of a discharge source based on a piezoelectric stack actuator by Suits and co-workers that allowed the use of much higher actuation forces than conventional piezo disks. \cite{abeysekera14a,shi15a} The backing pressures used in these pulsed fluorine discharge sources were relatively low. Benefits of higher backing pressures, such as increased beam densities, narrower velocity spreads and internally colder beams, were achieved by combining Even-Lavie-type gas valves with a DBD or a conventional plate discharge resulting in atomic beams of a wide range of different radicals.\cite{even00a,luria09a,lu07a} 

Here, we present a corrosion-resistant atomic fluorine source based on a pulsed solenoid valve which can be operated at high backing pressures and high operation frequencies. The use of corrosion-resistant materials, such as stainless steel, macor and teflon, enables stable operation for months with little to no maintenance. We compare and characterize in detail two different dissociation techniques, a conventional plate discharge and a dielectric-barrier discharge, both of which are often used in crossed-beam experiments for the production of radicals, metastable species or ions. 

The high ionization energy of atomic fluorine of 17.4~eV poses challenges for the characterization of fluorine radical sources. In the past, the direct detection of F involved either electron paramagnetic resonance spectroscopy\cite{carrington66a} or the generation of vacuum-ultraviolet (VUV) radiation for single- or multi-photon ionization detection schemes,\cite{bemand73a,herring88a,ren13a} both of which were associated with challenging requirements for the experimental setup. Incomplete dissociation of molecular fluorine in the discharge requires the ability to differentiate between fluorine originating from the discharge process and from spurious photodissociation by the probe laser for an accurate characterization of the radical beam. Various techniques have previously been used for the indirect detection of fluorine atoms, e.g., by titration with Cl$_{2}$,\cite{habdas89a,clyne73a} by elastic scattering from an inert gas followed by electron bombardment\cite{faubel94a,rusin06a} or by probing the products of reactive collisions.\cite{ren06a} These rather slow detection methods are, however, not ideal in the context of optimization procedures requiring a fast response from the experiment. Here, we present the adaption of a (3+1) resonance-enhanced multiphoton ionization (REMPI) scheme employed by Roth et al.\cite{roth99a} and universal femtosecond-laser ionization in combination with time-of-flight mass spectrometry and velocity-map imaging\cite{eppink97a} as a tool for the optimization and characterization of the present fluorine-radical sources. 

\section{Experimental setup}

\begin{figure} [h]
    \centering
    \includegraphics[scale=0.195]{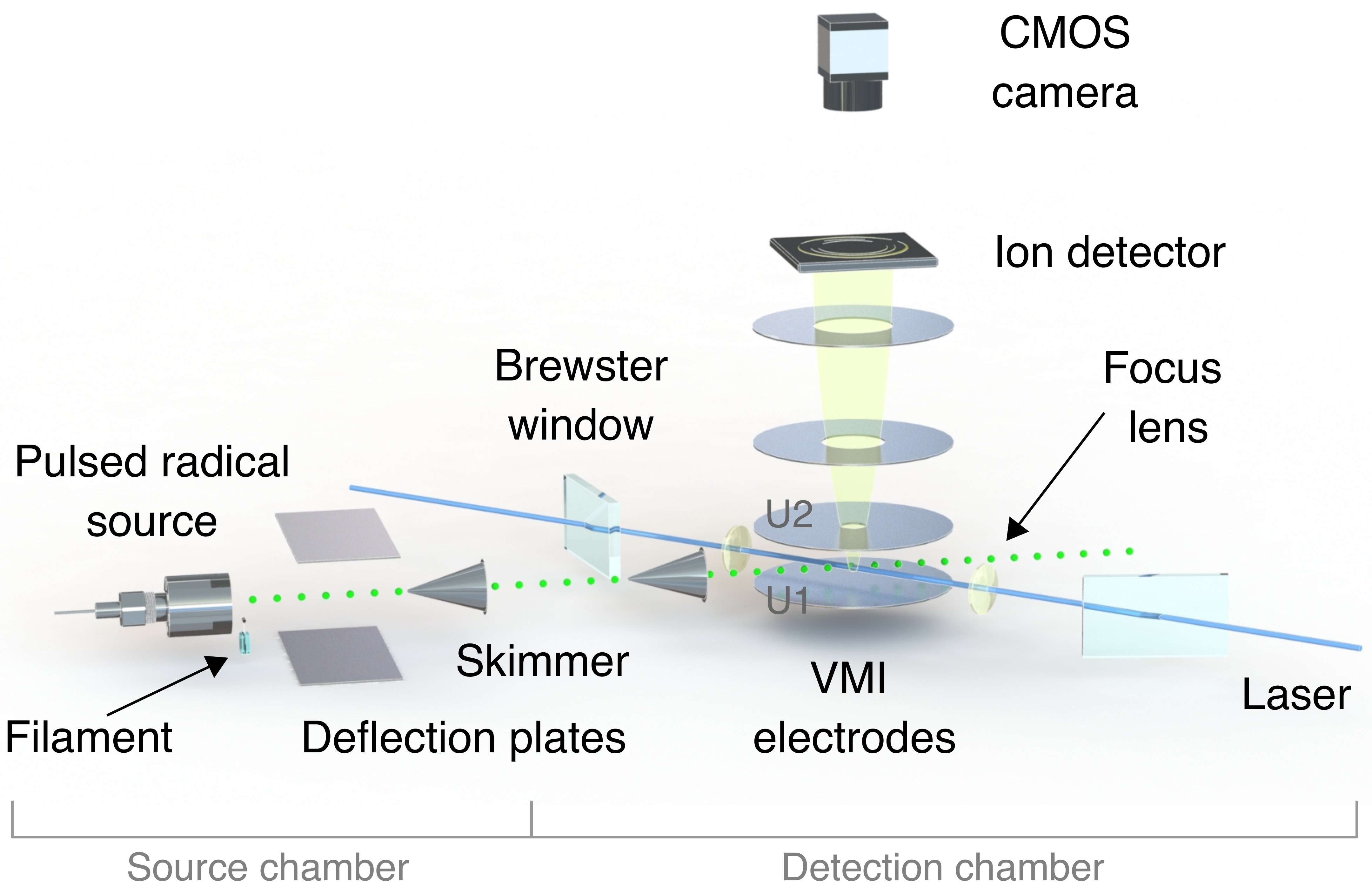}
    \caption{Experimental setup used for the characterization of the present discharge sources. See text for details. }
    \label{fig:expsetup} 
\end{figure}

The discharge sources were characterized and optimized using a CMB apparatus recently set up in our laboratory.\cite{ploenes21} A schematic of the part of the apparatus relevant for the current experiments is depicted in Fig.~\ref{fig:expsetup}. It consisted of two differentially pumped vacuum chambers separated by a conical skimmer. The source and the detection chambers were pumped by two and one turbo-molecular pumps (Oerlikon Leybold MAG W 2200), respectively, with pumping speeds of 2200 ls$^{-1}$ each. A first beam skimmer (orifice diameter 2 mm) was placed 11 cm downstream from the discharge source nozzle. A second skimmer with an orifice diameter of 1.5~mm was placed in front of the detection region to further confine the beam. Electrostatic deflection plates were placed in between the nozzle and the first skimmer in order to clean the beam from spurious ions produced in the discharge. \cite{luria09a} The total distance from the nozzle to the intersection region with the probe laser amounted to 46 cm. 

The discharge units were attached to a solenoid-based pulsed valve designed in the Canadian Center for Research on Ultra-Cold Systems (CRUCS).\cite{grzesiak18a} The valve was mounted on an $xyz$-translational stage inside the source chamber. The duration for opening the valve during a gas pulse was set to be nominally $32~\mu$s identical for both discharge sources to enable a direct comparison. Typical operating pressures in the source and detection chamber at 20 Hz repetition rate were in the low 10$^{-6}$ and 10$^{-8}$ mbar regime, respectively. An increase of the repetition rate of the experiment to 200 Hz increased the pressures in both chambers by a factor of 10.
A molecular beam (MB) of fluorine was formed by a supersonic expansion of a commercially available mixture of 5\% F$_{2}$ in He (Carbagas AG). The F radicals were created by dissociation of the F$_2$ molecules using the electric-discharge sources. 

\begin{figure} [h]
    \centering
    \includegraphics[scale=0.305]{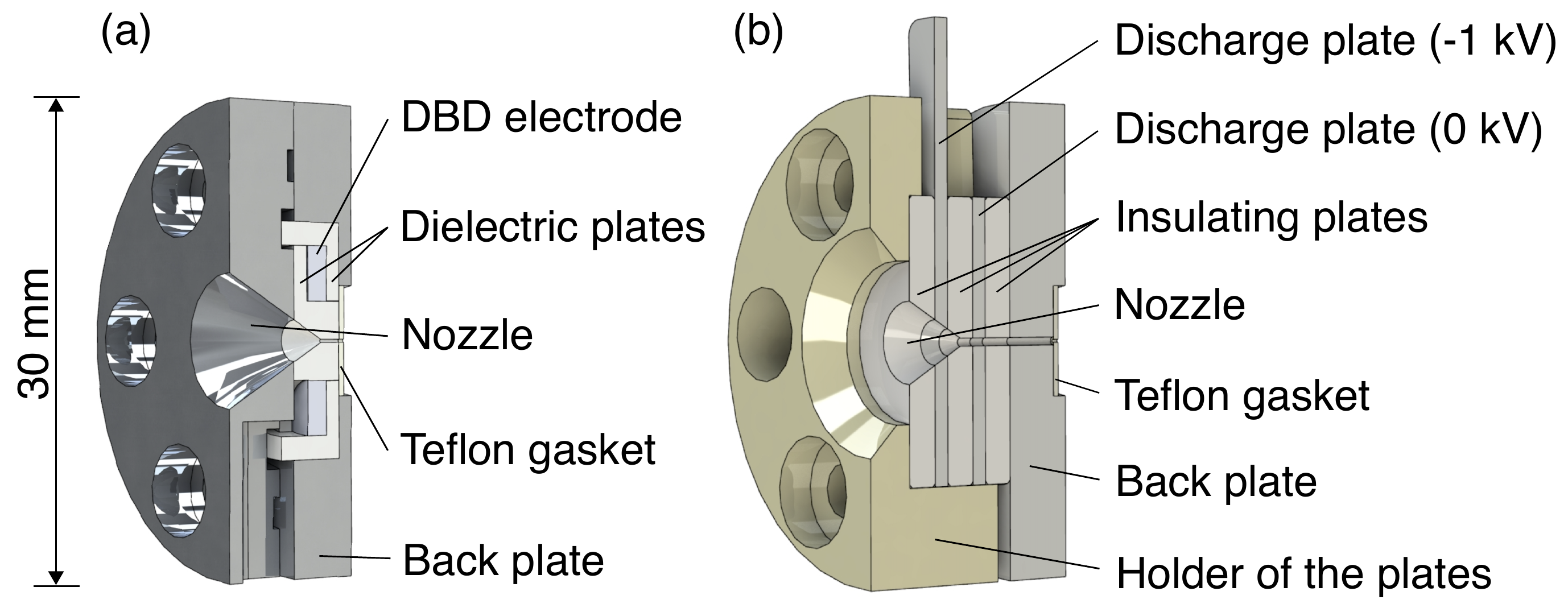}
    \caption{
    (a) Schematic illustration of the dielectric-barrier-discharge (DBD) unit. The diameter of the nozzle channel amounted to 0.2 mm. The thickness of the DBD electrode was 1.5 mm and its outer diameter was 12 mm. Dielectric plates were fabricated from Macor\textsuperscript{\textregistered}. The nozzle exit consisted of a cone with an opening angle of 60$^{\circ}$. (b) Schematic illustration of the plate-discharge source. The discharge plates (thickness 1~mm, outer diameter 18~mm) were separated by 2~mm thick insulators. The gas channel with an inner diameter of 0.5~mm opens into a conically shaped nozzle with an opening angle of 60$^{\circ}$ starting at the central insulating plate. Sealing between the plunger of the valve and both discharge units was achieved by a 0.4 mm thick, corrosion-resistant Teflon gasket (inner diameter 0.2~mm)
    }
    \label{fig:DBD_Plate_design} 
\end{figure}

The design of the discharge units was based on sources originally developed for the generation of dense beams of OH radicals. \cite{ploenes16a} Construction details of the sources are depicted in Fig.~\ref{fig:DBD_Plate_design}. For the DBD source [Fig.~\ref{fig:DBD_Plate_design} (a)], the electric discharge takes place at the surface of a dielectric material (Macor\textsuperscript{\textregistered}) in the form of microdischarge channels between the inner electrode (a magnetic disc) and the valve body. \cite{brandenburg17} The DBD is generated by applying an AC voltage to the electrode, with a peak-to-peak amplitude of 4.5 kV and a frequency of 780 kHz. The amplitude was limited by the commercial driver used (Lamid Ltd.). The optimal delay of the discharge pulse relative to the valve opening trigger was found to be $30~\mu$s.

For the plate-discharge source [Fig.~\ref{fig:DBD_Plate_design}(b)], the electric discharge is created between two metal plates isolated by electric insulators (Macor\textsuperscript{\textregistered}) by applying a voltage of -1kV to the electrode intersecting the nozzle orifice. A home-built high-voltage driver was used for igniting the discharge. Because the distance between the electrodes and the valve orifice is increased compared to the DBD source, the optimum delay between the gas and discharge pulses had to be raised to $55~\mu$s compared to the DBD source.

The addition of an electron-emitting filament in front of the nozzle, as reported in Ref.~\onlinecite{lewandowski04a}, increased the  stability of the DBD source. This was not required for the plate discharge. Since the filament was also introducing additional heat into the system, it was nonetheless kept operational also for the plate discharge in order to maintain the setup at the same temperature and thus provide the same conditions for a direct comparison between the discharge sources. 

Fluorine radicals were state-selectively ionized using a (3+1) REMPI scheme following Ref.~\onlinecite{roth99a}. The output of a pulsed dye laser (Liop-Tec LiopStar E) pumped by a Nd:YAG laser (Innolas Spitlight 1000) at a repetition rate of 20 Hz was frequency doubled by a $\beta$-barium borate (BBO) crystal to produce radiation at a wavelength of 234~nm. The laser beam entered and exited the chamber via two Brewster windows in order to minimize stray light. The light was focused into the interaction region by a lens with a focal length of 70~mm placed inside the vacuum chamber. The polarization of the light was chosen parallel to the imaging plane. Measurements at higher repetition rates up to 200~Hz were performed using universal 775~nm ionization with a femtosecond laser (Clark MXR) with a pulse duration of 150 fs.

The detection of the ionization products was performed by time-of-flight mass spectrometry (TOF-MS) and mass-specific velocity-map imaging (VMI).\cite{eppink97a} The ions were accelerated towards a double-stack microchannel-plates (MCP) detector connected to a phosphor (P46) screen (Photek, Ltd). In VMI mode, mass selection was achieved by time-gating the MCPs. In TOF-MS mode, the electrodes were operated in Wiley-McLaren configuration\cite{wiley55a} by applying $U_1$ = 3000~V and $U_2$ = 2000~V to the extraction plates. To achieve optimal velocity focusing in VMI mode, $U_2$ was increased to 2235 V.

Both discharge sources were operated for two hours before taking any measurements to reach stable experimental conditions. Additionally, to ensure a fair comparison of both discharge mechanisms, the density of the neat molecular beam prior to dissociation was verified to be constant in all experiments by monitoring the number of F ions produced by the (3+1) REMPI ionization of F$_2$.


\section{Results and discussion}

\subsection{(3+1) REMPI spectrum of F atoms}

As a preparation for the characterization and optimization of the discharge sources, (3+1) REMPI spectra of fluorine atoms originating from the photodissociation of F$_{2}$ in the absence of the discharge were recorded. At 234~nm, photodissociation of F$_{2}$
\begin{equation} \label{eq1}
\text{F$_{2}$} + hv \rightarrow \text{2F($^{2}$P$_{J}$)},
\end{equation}
occurs via the repulsive state A$^{1}\Pi_\text{u}$.\cite{roth99a} Fluorine atoms generated in different spin-orbit components $^{2}$P$_{J}$, where $J$ denotes the quantum number of the total atomic angular momentum excluding nuclear spin, were further ionized by (3+1) REMPI using the same wavelength and the resulting ion signal was detected using TOF-MS.

\begin{figure} [t]
    \centering
    \includegraphics[scale=0.47]{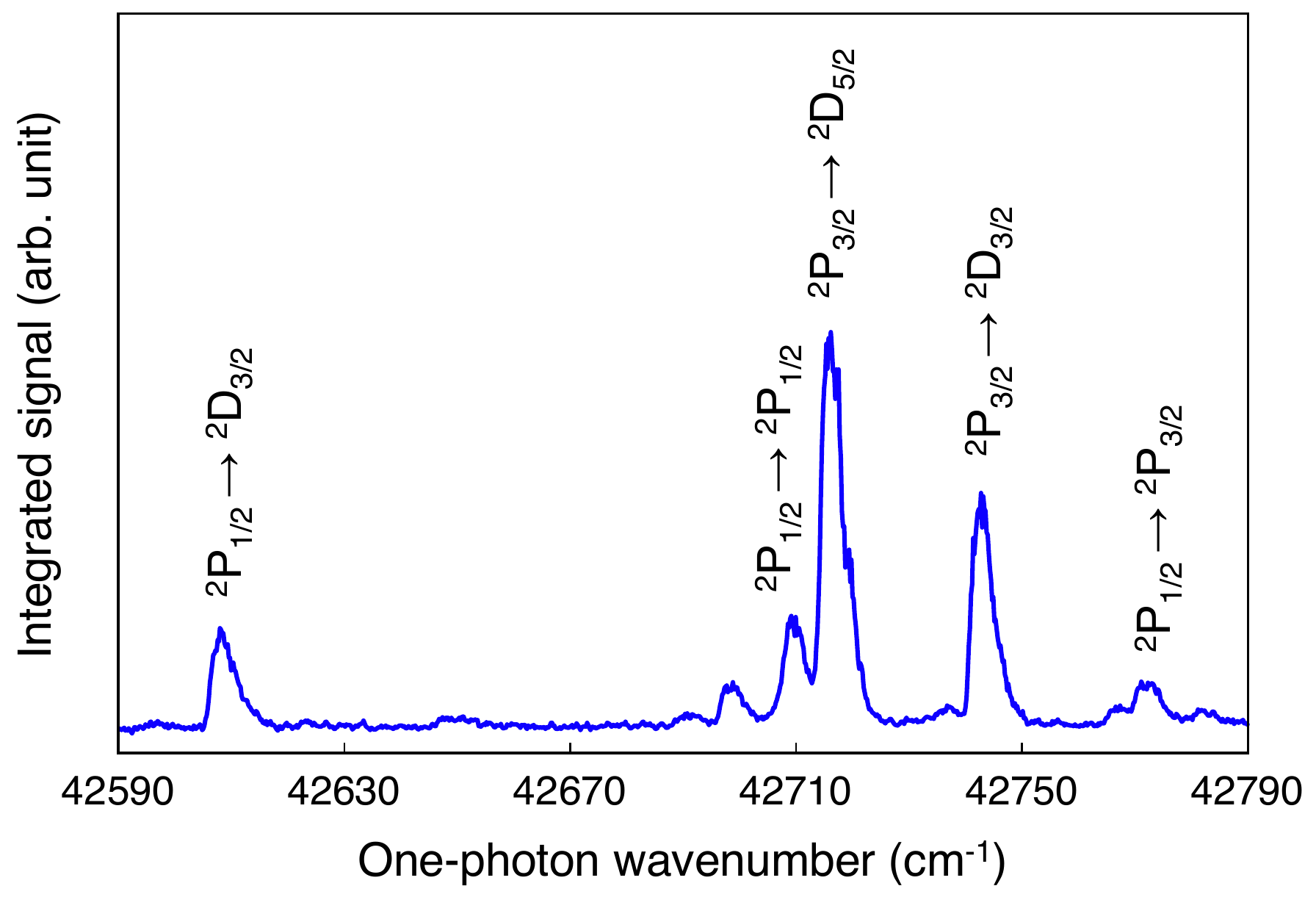}
    \caption{(3+1) resonance-enhanced multiphoton ionization (REMPI) spectrum of fluorine atoms produced by the photodissociation of F$_{2}$ by the REMPI laser. The excitation of the atoms is carried out via the (2p$^{5}$) $^{2}$P$_{3/2}/^{2}$P$_{1/2} \rightarrow (2p^{4}3$d$) ^{2}$D$_{j}/^{2}$P$_{k}$ transitions as indicated.}
    \label{fig:rempi_scan} 
\end{figure}

An example of a REMPI spectrum is shown in Fig.~\ref{fig:rempi_scan}. The one-photon wavenumber was scanned from 42590 to 42790 cm$^{-1}$ at a pulse energy of 0.3 mJ in the interaction region. The spectrum exhibits transitions from the ground (2p$^{5}$) $^{2}$P$_{3/2}$ and spin-orbit excited (2p$^{5}$) $^{2}$P$_{1/2}$ states to the (2p$^{4}$3d) $^{2}$D$_{j}$ and $^{2}$P$_{k}$ levels by the non-resonant absorption of three photons. A slight systematic shift of $\approx 3$~cm$^{-1}$ of the transition wavenumbers in Fig.~\ref{fig:rempi_scan} in comparison to the literature values \cite{roth99a} is attributed to differences in the wavelength calibration (here, a HighFinesse WS-6 wavemeter with an internal Ne calibration lamp  was used) and Doppler shifts due to the oblique positioning of the laser with respect to the molecular beam.

\subsection{Comparison of discharge sources using velocity-map imaging}

\begin{figure*}[!ht]
    \centering
    \includegraphics[width=\textwidth]{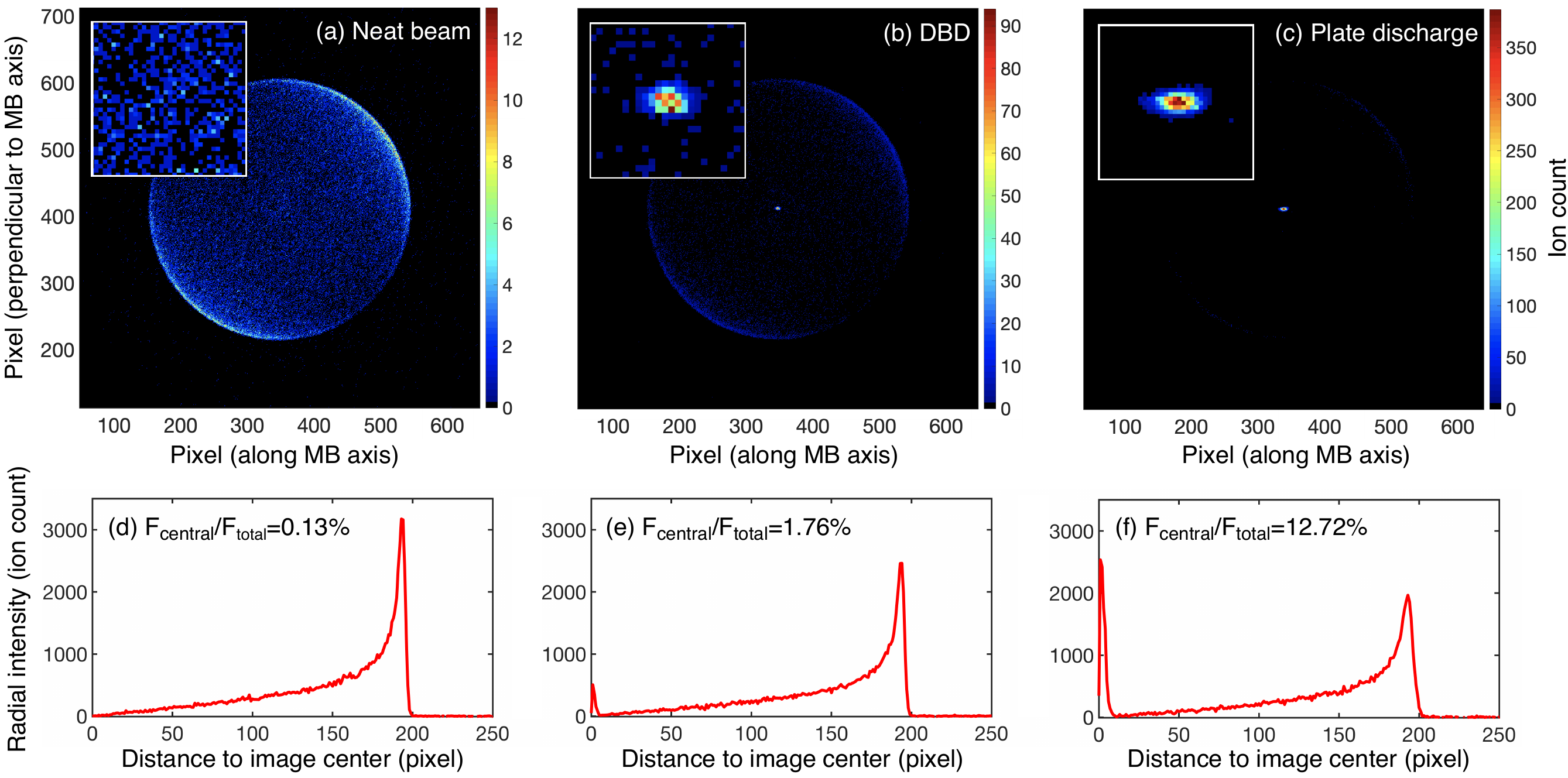}
    \caption{Velocity-mapped ion images (top row) of fluorine atoms ionized by (3+1) REMPI recorded at an experimental repetition rate of 20 Hz and the corresponding radial product distributions (bottom row). The VMI images were accumulated over 75 000 shots at the laser pulse energy of 0.15 mJ on the $^{2}$P$_{3/2}\rightarrow^{2}$D$_{5/2}$ (3+1) REMPI transition. The insets show magnifications of the central part of the image within an area of 24x24 pixels where the discharge products can be observed. Note the different intensity scales in the images. (a,d) Discharge switched off (valve backing pressure 10 bar); (b,e) DBD source (10 bar); (c,f)  plate-discharge source (30 bar).}
    \label{fig:dbd_vs_plate} 
\end{figure*}

The discharge sources were optimized by measuring the yield of fluorine radicals on the $^{2}$P$_{3/2}\rightarrow^{2}$D$_{5/2}$ (3+1) REMPI transition (Fig.~\ref{fig:rempi_scan}) using VMI. The velocity-resolved images allowed us to distinguish F radicals produced in the discharge source from atoms generated by the photodissociation of F$_{2}$ by the REMPI laser. F atoms arising from the photodissociation process are generated with high recoil velocities and thus appear as rings in the VMI images. By contrast, F radicals produced by the discharge possess near zero recoil velocity after laser ionization and thus appear as a spot in the center of the VMI images. Exemplary results of these measurements are shown in Figs. \ref{fig:dbd_vs_plate}(a)-\ref{fig:dbd_vs_plate}(c) which compare VMI images obtained from a neat F$_2$/He beam with ones obtained using the DBD and plate discharges. 

Integration of the VMI images over the angular coordinate resulted in the radial product distributions shown in Figs. \ref{fig:dbd_vs_plate}(d)-\ref{fig:dbd_vs_plate}(f). In contrast to the neat beam [Fig.~\ref{fig:dbd_vs_plate}(a) and \ref{fig:dbd_vs_plate}(d)], for both the DBD- and plate-discharged beam a clear signal was observed in the center of the images which is attributed to the fluorine radicals generated by the discharges. This is accompanied by a decrease in fluorine signal originating from the photodissociation of F$_2$.

Both discharge sources were measured under their respective optimized conditions established by maximizing the fluorine signal accumulated in the centre of the images. For the DBD source, the maximum F yield was reached at a valve backing pressure of 10~bar, whereas for the plate-discharge source, the highest yields were achieved at a backing pressure of 30~bar. A further increase of the backing pressure was not attempted due to limitations of the  corrosion-resistant pressure regulator used. The discharge efficiencies of the DBD and the plate discharges were compared by integrating the fluorine signal in the center of the images, while taking into account a small background contribution. Based on this comparison, the radical yield of the plate discharge was estimated to be 8-9 times higher than that of the DBD. Using the same backing pressure of 10~bar for both sources, the plate discharge still produces 2-3 times higher fluorine atom fluxes. Alternatively, a comparison of the ratios of fluorine radicals produced by the discharge and by photodissociation, as shown in the radial product distributions (d), (e) and (f) of Fig.~\ref{fig:dbd_vs_plate}, gives a factor of 7.2 in radical yield between the plate-discharge and the DBD sources which is similar to the result obtained from comparing the integrated signals of the central spots. These results indicate a markedly higher efficiency of the plate discharge compared to the DBD for producing F atoms. 

\subsection{Relative state populations of F radicals generated by the two methods}

\begin{figure} [h]
    \centering
    \includegraphics[scale=0.47]{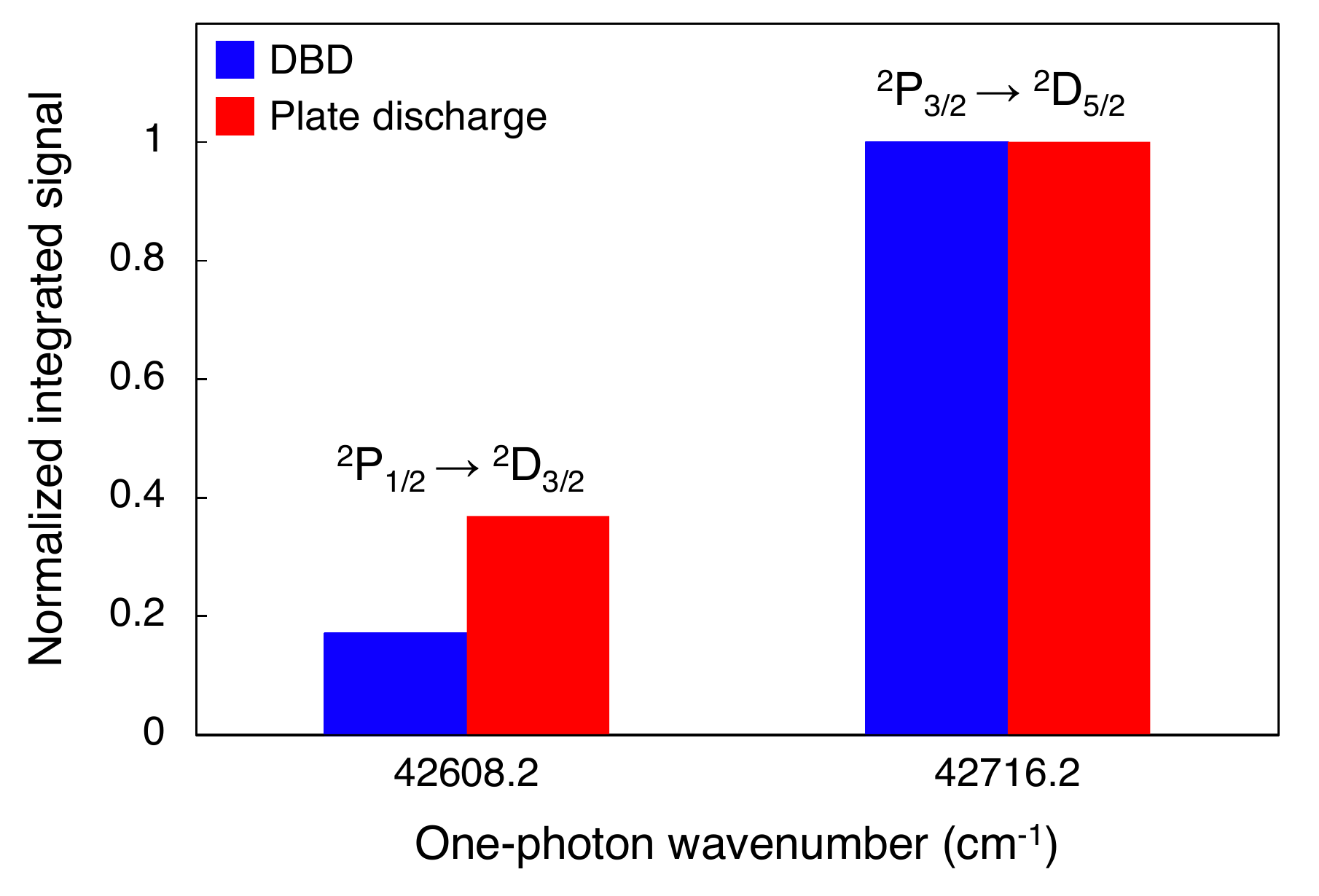}
    \caption{Comparison of the relative intensities of selected (3+1) REMPI transitions (see Fig. \ref{fig:rempi_scan}) originating from the ground and spin-orbit excited states of F produced by the DBD and plate-discharge sources. The relative strengths of the transitions indicate the relative populations of the two spin-orbit components. All intensities were normalized with respect to the $^{2}$P$_{3/2}\rightarrow^{2}$D$_{5/2}$ transition.} 
    \label{fig:radbeamtemp} 
\end{figure}

By comparing the intensity of different transitions in the (3+1) REMPI spectra of the F atoms generated in the discharge sources, the relative populations of the ground $^{2}$P$_{3/2}$ and spin-orbit excited $^{2}$P$_{1/2}$ states were determined by integrating the relevant signal peaks for the DBD and plate discharge (Fig.~\ref{fig:radbeamtemp}). The plate-discharge source produces more radicals in the spin-orbit excited $^{2}$P$_{1/2}$ state, indicating an internally hotter radical beam compared to the DBD source. Similar results were previously obtained for the generation of OH radicals with both discharge techniques.\cite{ploenes16a} 

\subsection{Operation of the sources at high experimental repetition rates}

\begin{figure}
    \centering
    \includegraphics[scale=0.385]{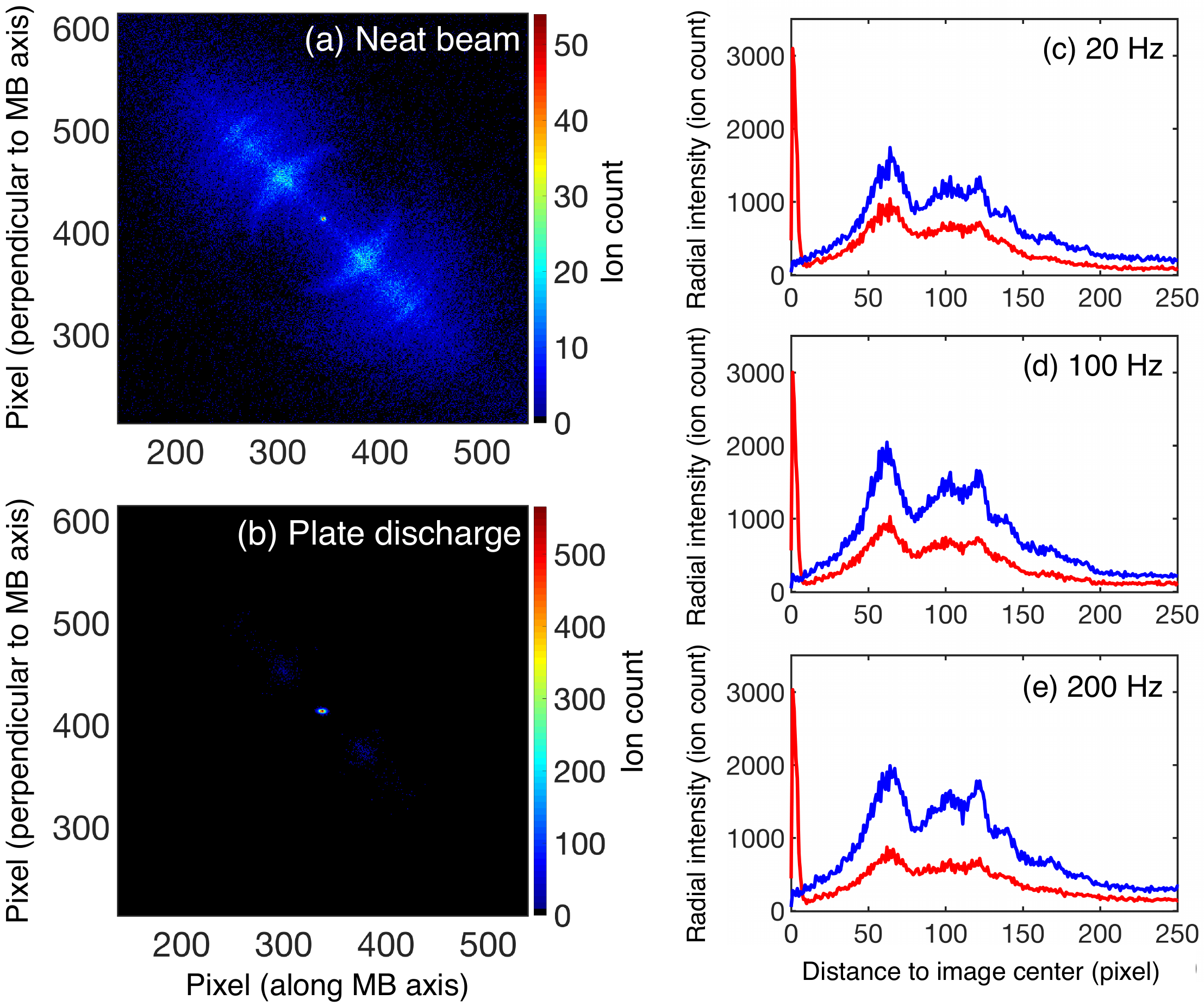}
    \caption{Examples of VMI images of F ions taken at an experimental repetition rate of 100 Hz using femtosecond-laser ionization of (a) a neat beam of F$_2$ seeded in He and (b) a beam subjected to the plate discharge. The enhancement of the F signal by the discharge is visible in the center of the image (note the different scales of the heat map in (a) and (b)). (c)-(e) Radial product distribution functions extracted from VMI images at repetition rates of 20, 100 and 200 Hz of a neat (blue traces) and a discharged (red traces) beam. No significant degradation of the performance of the discharge source could be observed across the range of repetition rates studied. The VMI images were averaged over 20 000 shots at the laser pulse energy of 0.12 mJ.}
    \label{fig:highfreq} 
\end{figure}

In order to test the performance of the plate-discharge source at various experimental repetition rates up to 200 Hz, VMI images obtained by femtosecond-laser ionization were recorded. Figure \ref{fig:highfreq} shows an example of VMI images of the (a) neat and (b) discharged beam at a repetition rate of 100 Hz. With fs-laser ionization, the VMI images appear highly structured containing a number of features. These are associated with fs-laser multiphoton dissociation of molecular fluorine producing F atoms in a variety of internal states.\cite{hijikata61} In contrast to (3+1) REMPI ionization, the multiphoton excitation and ionization by the fs laser partly also produces slow F$^+$ ions which appear in the center of the image and mask the contribution of F atoms generated by the discharge, as can be seen in Fig.~\ref{fig:highfreq}(a).

Radial product distribution functions obtained from the VMI images taken at 20, 100 and 200 Hz [Figs. \ref{fig:highfreq}(c)-\ref{fig:highfreq}(e)] show no significant difference in the performance of the discharge source as a function of the repetition rate. Achieving even higher repetition rates was only prevented by the increased heating of the valve caused by the driving current of the valve mechanism and the plunger friction. Additional cooling of the valve should enable to overcome this limitation and reach even higher repetition rates. A similar constant performance as a function of the repetition rate was observed for the DBD source (not shown here). Analysis of the ion counts in these images indicates that the plate-discharge source produces about a factor of 7 more F radicals than the DBD source at each repetition rate studied, in good agreement with the results obtained by (3+1) REMPI.

\begin{figure} [t]
    \centering
    \includegraphics[scale=0.305]{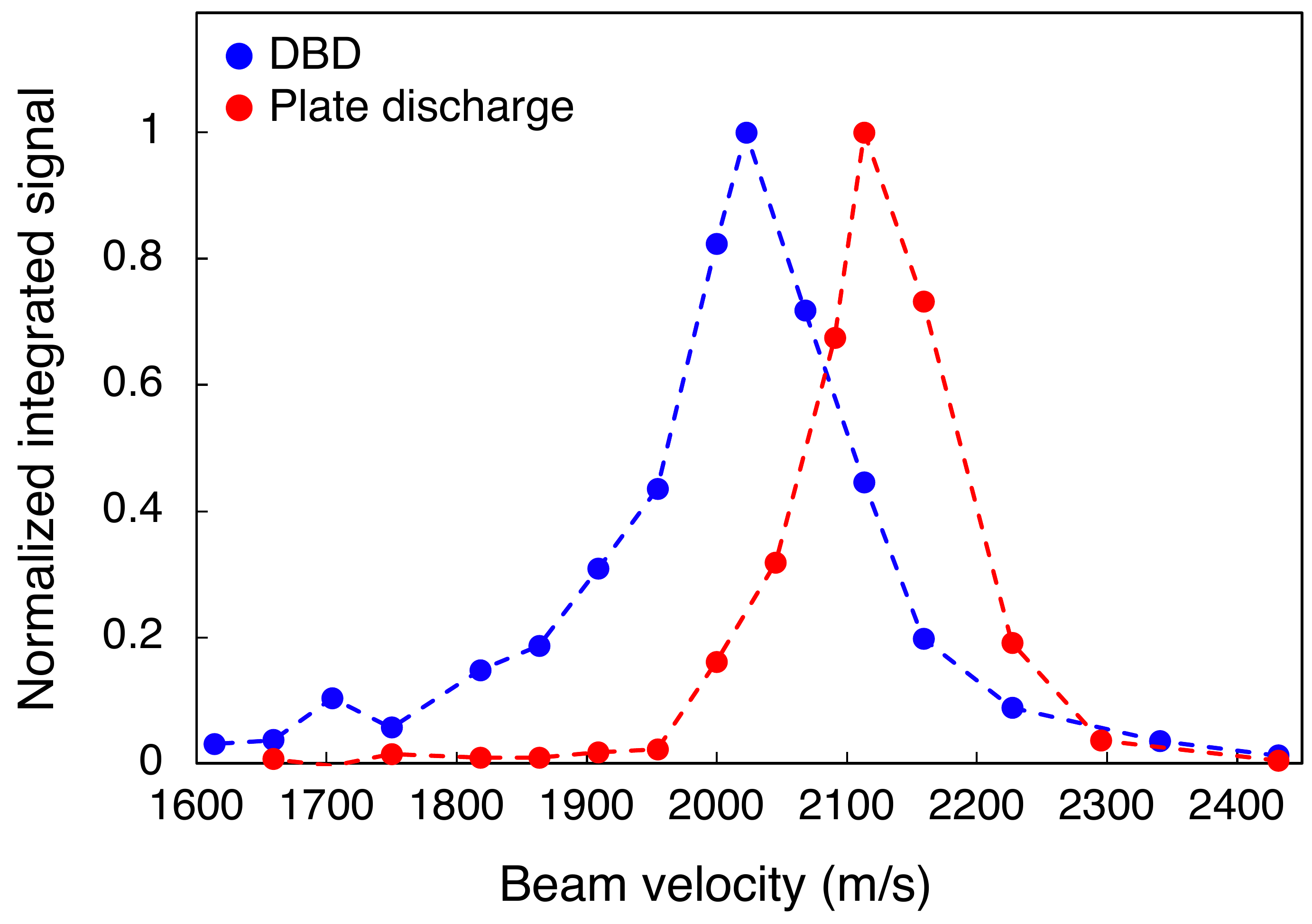}
    \caption{Typical fluorine radical beam profiles as a function of the mean beam velocity for the DBD and plate-discharge sources taken at a repetition rate of 100 Hz using fs-laser ionization.}
    \label{fig:profiles} 
\end{figure}

\subsection{Velocity profiles of the radical beams}

To obtain velocity profiles of the fluorine radical pulses, VMI images were measured at a repetition rate of 100 Hz for different time delays of the gas beam with respect to the trigger of the fs laser. Velocities of the beam at each time delay were extracted from these measurements. Normalized velocity profiles of the radical beam profiles after subtracting the contribution from photodissociated fluorine atoms are shown in Fig.~\ref{fig:profiles} for both discharge sources. The profiles are very similar. The speed ratio estimated from the velocity at maximum intensity with $\Delta v$, the full width at half maximum, was determined to be $v/\Delta v\approx 14$ and 18 for the DBD and plate-discharge source, respectively. A slightly faster velocity was observed for the beam produced by the plate discharge which is attributed to the more violent character of this type of discharge.


\section{Summary and conclusions}
We have presented two corrosion-resistant discharge sources, a high-voltage plate discharge and a dielectric-barrier discharge, for the generation of beams of atomic fluorine radicals operational at repetition rates up to 200 Hz and high backing pressures. Their performance was analyzed with the help of velocity-map ion imaging of the generated F atoms and compared in terms of their efficiencies and beam properties. The combination of the discharge sources with a high-pressure, short-pulse solenoid valve produced high-density atomic beams of comparable narrow longitudinal velocity spreads. It was found that the plate discharge produces denser fluorine radical beams at the cost of increased internal temperature, as has also been observed in the production of OH radicals. \cite{ploenes16a} Both discharge sources exhibit an excellent resistivity to corrosion thus offering long-term stable operation without the requirement of frequent maintenance. The sources were intensively used for the production of fluorine radicals over several months with only requiring the exchange of teflon gaskets every few weeks. Both sources showed no dependence of their discharge efficiency on the experimental repetition rate up to 200 Hz making them suitable for experiments running at high frequencies. 

We note that the present plate-discharge source was also recently successfully applied in a study of the chemi-ionization reaction of metastable neon atoms with carbonyl sulfide molecules.\cite{ploenes21}

\begin{acknowledgments}
We thank Philipp Knöpfel, Grischa Martin, Georg Holderied and Anatoly Johnson for technical support. This work was supported by the Swiss National Science Foundation under grant nr. BSCGI0\_15787 and the University of Basel.
\end{acknowledgments}

\section*{Data availability}

The data that support the findings of this study are available from the corresponding author upon reasonable request.

\end{document}